\documentclass[aps,twocolumn,groupedaddress,amsfont,showpacs]{revtex4}
\usepackage{amssymb}

\usepackage[dvips]{graphicx}

\begin{document}
\preprint{Preprint \today}
\title{Bifurcations and chaos in semiconductor superlattices with a tilted magnetic
field.}

\author{A.\,G.~Balanov}
\author{D. Fowler}
\author{A. Patan\`e}
\author{L. Eaves}
\author{T.M. Fromhold}

\affiliation{School of Physics and Astronomy, University of Nottingham, Nottingham NG7 2RD, UK}

\begin{abstract}
We study the effects of dissipation on electron transport in a semiconductor
superlattice with an applied bias voltage and a magnetic field that is tilted relative to the superlattice axis.
In previous work, we showed that although the applied fields are stationary,
they act like a THz plane wave, which strongly couples the Bloch and
cyclotron motion of electrons within the lowest miniband. As a consequence,
the electrons exhibit a unique type of Hamiltonian chaos, which creates an intricate mesh of conduction channels (a stochastic web) in phase space, leading to a large resonant increase in the current flow at critical values of the applied
voltage. This phase-space patterning provides a sensitive mechanism for controlling electrical resistance. In this paper, we investigate the effects of dissipation on the electron dynamics by modifying the semiclassical equations of motion to include a linear
damping term. We demonstrate that even in the presence of dissipation,
deterministic chaos plays an important role in the electron transport process. We
identify mechanisms for the onset of chaos and explore the associated
sequence of bifurcations in the electron trajectories. When the Bloch and cyclotron frequencies are
commensurate, complex multistability phenomena occur in the system. In
particular, for fixed values of the control parameters several distinct
stable regimes can coexist, each corresponding to different initial conditions. We show that this multistability has clear, experimentally-observable, signatures in the electron transport characteristics.

\end{abstract}


\maketitle
\section{Introduction.}

\begin{figure}[t!]
  \centering
  \includegraphics[width=1.0\linewidth]{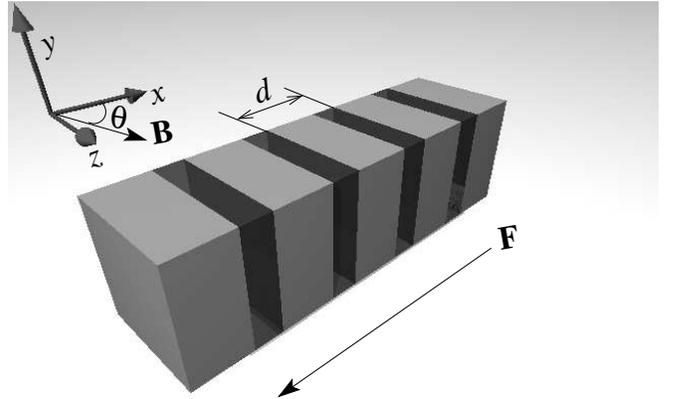}
  \caption{Schematic layer structure of a semiconductor SL formed from two different semiconductor materials, shaded light and dark gray. The co-ordinate axes show the orientation of the tilted magnetic field, $\mathbf{B}$, which lies in the $x-z$ plane at an angle $\theta$ to the SL ($x$) axis. The lower arrow shows direction of the electric field, $\mathbf{F}$, applied perpendicular to the plane of the layers and anti-parallel to the $x$ axis.}
  \label{schema}
\end{figure}

Semiconductor superlattices (SLs) are nanostructures made from alternating
layers of two different semiconductor materials, usually with very similar lattice constants, for example GaAs and (AlGa)As \cite{WAC02}. Due to the different energy band gaps of the two materials, the conduction band edge of an ideal superlattice is
periodically modulated. 
Typically, a SL made by molecular beam epitaxy contains 10-100 quantum wells in series, which are coupled
by tunnel barriers. This periodic potential leads to the formation of energy
bands, known as ``minibands'', for electron motion perpendicular to the
layers \cite{WAC02}. 
A voltage applied to ohmic contacts at the two ends of the SL generates an electric
field, $\mathbf{F}$, perpendicular to the plane of the layers (see Fig. \ref
{schema}), which causes charge to flow through the device. The current-voltage
characteristics of SLs are usually highly nonlinear due to a
variety of quantum mechanical effects including resonant tunneling, the
formation of Wannier-Stark energy level ladders, and the occurrence of Bloch
oscillations \cite{SHI75,WAC02}, whose frequency is proportional to the spatial period, $d$,  of the SL and, also, to $F$. In natural crystals, 
$d$ is so small ($\sim0.3$ nm) that Bloch
oscillations do not occur because the Bloch frequency is much less than the
electron scattering rate. But in SLs, $d$ and the corresponding Bloch frequency can be
large enough for the electrons to perform Bloch oscillations, which then
play a key role in both the dc and high-frequency charge transport processes. The onset of Bloch oscillations
localizes the electrons, thus causing their drift (average) velocity to
decrease as the electric field increases. This negative differential
velocity can induce high-frequency collective oscillations of the
conduction electrons within the SL layers \cite{ESA70}, making SLs attractive for the generation
and detection of electromagnetic radiation in the GHz to THz frequency range \cite{SCH98,RAS05,SHI03,SAV04}. 

The occurrence of negative differential velocity for electrons in SLs 
has also led to fundamentally new regimes of charge transport involving deterministic chaos, characterized by complex
irregular electron dynamics \cite{WAC02,SCH01,BON05}. Understanding the effects of
chaos on the current-voltage characteristics, $I(V)$, and high-frequency
electromagnetic properties of SLs is an emerging research area at the
interface between nonlinear dynamics and condensed matter physics.
In most previous studies, chaos in SLs was identified in the cooperative motion of
interacting electrons in either periodically driven \cite{ZHA96, ALE96} or
undriven SLs \cite{AMA02a}.

Recent theoretical and experimental work has revealed that the single-particle trajectories and collective behavior of electrons moving through the lowest miniband of a biased SL with a tilted magnetic field have unique properties \cite{FRO01,FRO04,Trav,KOS06}.
For example, the electrons exhibit 
 an unusual type of
Hamiltonian chaos, known as ``non-KAM'' chaos \cite{SAG88,ZAS91,VAS88}, which does not obey the Kolmogorov-Arnold-Moser (KAM)
theorem and provides a sensitive new mechanism for controlling electron
transport.
Remarkably, the effective classical Hamiltonian for the electron
motion in a SL has an intrinsically quantum-mechanical origin as it depends
explicitly on the energy versus wavevector dispersion relation of the
miniband \cite{FRO01,FRO04}. Analysis of Hamilton's equations reveals that
the stochastic electron motion switches on abruptly when the field parameters satisfy
certain resonance conditions, described in detail below. The onset of chaos
delocalizes the electrons by imprinting an intricate mesh of conduction
channels (known as a ``stochastic web'') in phase space, which produces
large resonant enhancement of the electron velocity and current flow
measured in experiment \cite{FRO04}. This phase-space patterning provides a
fundamentally new concept for controlling electrical conductance \cite{STA04} and
operates even at room temperature \cite{FRO04}.

In the absence of dissipation, the resonant delocalization of the conduction electrons would produce a series of
delta-function peaks in the $I(V)$\ curves. But in real SLs, the resonances are broadened because the electrons
scatter both elastically \cite{WAC02}, due mainly to roughness at the
interfaces between the tunnel barriers and quantum wells and to ionized donor atoms, and inelastically via the emission
and absorption of phonons. When calculating
the transport characteristics of the SLs, the effect of scattering on the
electron dynamics must be taken into account in order to obtain $I(V)$\
curves that agree with experiment. In our
previous theoretical work \cite{FRO01,FRO04}, we incorporated electron scattering in the
following phenomenological way. First, we calculated the electron
trajectories by solving Hamilton's equations for the system, assuming $no$
scattering. Next, we used these trajectories to determine the average
electron velocity from a simple kinetic formula in which scattering appears
{\it a posteriori} as an exponential damping term whose decay rate depends on the
elastic and inelastic relaxation times. In this paper, we investigate
whether the striking electron resonance effects and phase space patterning
predicted by our earlier collision-free Hamiltonian model of the electron
dynamics persist when scattering is included {\it a priori} in the equations of
motion. We find that when the scattering is described in this way, it
creates far richer electron dynamics than expected from our previous
Hamiltonian analysis. In particular, scattering makes the system highly sensitive to changes in the
control parameters (electric and magnetic fields), which induce complex
bifurcation sequences characterized by alternating windows of stability and
chaos. Counter-intuitively, dissipation also enhances the resonant
delocalization of the semiclassical trajectories by creating attractors that drive the
electrons rapidly through the SL. The regimes of dissipative chaos that we have identified have a pronounced effect on the
velocity-field characteristics of the miniband electrons and provide mechanisms for controlling SL magnetotransport by exploiting complex nonlinear dynamics.

\bigskip The structure of the paper is as follows. In Section II, we
introduce the semiclassical equations of motion for a miniband electron in a
tilted magnetic field and show how scattering is included \emph{a priori} in those
equations. In Section III, we show that scattering dramatically
changes the electron trajectories and phase space structure of the system by
creating attractors corresponding to periodic orbits that extend through the
whole SL. In Section IV, we present a detailed analysis of the stability of
the electron orbits. Section V explores the complex series of resonances that
scattering induces in the electron dynamics and elucidates the effect of these resonances on the
electron velocity and experimentally-measured electrical current. Finally, in Section VI, we summarize our results and draw
conclusions.

\section{Model equations.}
\label{sec:model}
In a SL, quantum mechanical tunneling broadens the discrete quantized energy levels of each individual quantum well into SL minibands. The miniband states are delocalized Bloch functions specified by the crystal momentum $\mathbf{p}=$ $(p_x,p_y,p_z)$=$\hbar\mathbf{k}$, where $\mathbf{k}$ is the corresponding electron wavevector. Within the tight-binding approximation, the energy versus crystal momentum dispersion relation for the lowest miniband is $E(\mathbf{p}%
)=\Delta[1-\cos(p_xd/\hbar)]/2+(p_y^2+p_z^2)/2m^{\ast}$, where $\Delta$ is the
miniband width, $d$ is the SL period, and $m^{\ast}$ is the electron effective
mass for motion in the $y-z$ plane. The crystal momentum component, $p_x$, is taken to lie within the first minizone of the SL.

Throughout this paper, we consider electron motion in an electric field $\mathbf{F}=(-F,0,0)$ applied anti-parallel to the $x$ axis, and a tilted magnetic field $\mathbf{B}%
=(B\cos\theta,0,B\sin\theta$) (Fig. \ref{schema}). In a semiclassical picture, which neglects inter-miniband tunneling, the force produced by the electric and magnetic fields changes the electron's crystal momentum at a rate 
\begin{eqnarray}
\frac{d\mathbf{p}}{dt}=-e(\mathbf{F}+(\nabla_\mathbf{p}E(\mathbf{p})\times\mathbf{B}),
\end{eqnarray}
where $e$ is the electronic charge. Equation (1) can be written in the component form
\begin{eqnarray}  \label{beq}
\dot{p}_x&=&eF-\bar{\omega}_cp_y\tan\theta\\
\dot{p}_y&=&\frac{d\Delta m^{\ast}\bar{\omega}_c}{2\hbar}\sin\left(\frac{p_x d}{\hbar}\right)\tan\theta-\bar{\omega}_cp_z \\
\dot{p}_z&=&\bar{\omega}_cp_y,
\end{eqnarray}
where the left hand terms are time derivatives of the crystal momentum components and $\bar{\omega}_c$=$eB\cos\theta/m^{\ast}$ is the cyclotron frequency corresponding to the magnetic field component along the $x$ axis. It follows from Eqs. (2-4) 
that 
\begin{eqnarray}  \label{disseq}
\ddot{p}_z+\bar{\omega}_c^2p_z=C\sin(K p_z-\omega_B t+\phi),
\end{eqnarray}
where $C=(-m^{\ast}\bar{\omega}_c^2d\Delta\tan\theta)/2\hbar$, $%
K=d\tan\theta/\hbar$, and $%
\omega_B=eFd/\hbar$ is the Bloch frequency. The phase, $\phi=d(p_x(t=0)+p_z(t=0)\tan\theta)/\hbar$, depends on the initial conditions and equals zero for electrons starting from rest \cite{FRO01,FRO04}.
Equation (5) describes completely the electron motion because its solution, $p_z(t)$, uniquely determines all of the other dynamical variables \cite{FRO04}. Consequently, to include the effects of dissipation {\it ab~initio} in the equations of motion, we formally introduce a relaxation term, $\alpha \dot{p}_z$, into the left hand side of Eq. (5), which becomes 
\begin{eqnarray}  \label{peneq}
\ddot{p}_z+\alpha\dot{p}_z+\bar{\omega}_c^2p_z=C\sin(K p_z-\omega_B t+\phi),
\end{eqnarray}
where $\alpha$ is the damping constant. In this approach, $\alpha$ has the same meaning as a constant collision rate in the Boltzmann equation. The dissipation term in Eq. (\ref{peneq}) directly affects the in-plane momentum component, $p_z$. But when $\theta \neq 0$, electron motion in the $x$, $y$, and $z$ directions is coupled, which means that the dissipation term affects all of the dynamical variables, thus simulating the scattering processes that occur in a real SL device. 

Recent experiments \cite{scatt1,fowler} have shown that the presence of a high magnetic field strongly affects the inelastic scattering mechanisms required for current to flow through a biased SL \cite{scatt2}. In particular, when $\theta=0$ and $\hbar\omega_c>\Delta$, the emission of longitudinal optical (LO) phonons is strongly suppressed if the phonon energy, $\hbar\omega_{LO}$ = 36 meV in GaAs, exceeds $\Delta$ \cite{scatt1,fowler}, which is the case for the SL that we studied in Ref. \cite{FRO04}. In this low-dissipation regime, the current falls almost to zero because the oscillatory Bloch motion along the $x$ direction is not damped by LO phonon emission. Our model equations (2-6) capture this magneto-suppression of the current, because Eqs. (2), (4), and (6) decouple in the limit $\theta\rightarrow0$, which means that the $x$ motion is undamped, and so the mean electron velocity is zero. In the experiments \cite{FRO04}, as $\theta$ increases from 0, the current rises dramatically because the selection rules that prevent LO phonon emission when $\theta=0$ are broken. Our model equations (2-6) describe this activation of LO phonon emission by increasing the coupling between motion parallel and perpendicular to $x$ as $\theta$ increases from 0, thus enhancing scattering-assisted transport through the SL. Consequently, the equations encapsulate the strongly $\theta$-dependent scattering rates observed experimentally in the high magnetic field regime that is the focus of this paper. As a consequence, however, they do not accurately describe the strongly damped Bloch motion that occurs in the limit $B=0$, which we do not consider here.

We used Eq. (\ref{peneq}) to investigate the effect of dissipation on the motion of electrons in SLs similar to those described and studied experimentally in \cite{FRO01,FRO04}, taking $\Delta=26.2$ meV, $%
d=10.3$ nm, and $m^{\ast}=0.067 m_e$, where $m_e$ is the mass of a free electron. In our analysis, we first solve Eq. (6) numerically to obtain $p_z(t)$, which we then use to determine the other crystal momentum components
\begin{eqnarray}
p_x&=&p_x(t=0)+eFt-(p_z-p_z(t=0))\tan\theta, \nonumber\\ p_y&=&\frac{\dot{p}_z}{\bar{\omega}_c},
\end{eqnarray}
and the electron velocity components
\begin{eqnarray}  \label{veleq}
\dot{x}=\frac{d\Delta}{2\hbar}\sin(K p_z-\omega_B t+\phi), ~ \dot{y}=\frac{\dot{%
p}_z}{\bar{\omega}_c m^{\ast}},~ \dot{z}=\frac{p_z}{m^{\ast}}.
\end{eqnarray}

Eq. (\ref{peneq}) describes the motion of a damped harmonic oscillator, whose natural frequency equals the cyclotron frequency $\bar{\omega}_c$, driven by a plane wave of wave number $K$ and frequency equal to the Bloch frequency $\omega_B$. The linear dissipation term describes the coupling of the oscillator to a ``reservoir" of many additional degrees of freedom, which, for the particular case of electrons in a SL, is provided by elastic and inelastic scattering processes.

An {\it undamped} harmonic oscillator driven by a plane wave is one of few systems known to exhibit non-KAM chaos \cite{SAG88,ZAS91}. Its rich dynamical properties have been studied by many authors and provide insights for understanding a wide range of problems in, for example, plasma physics, tokamak fusion, turbulent fluid dynamics, ion traps and quasi-crystals \cite{SAG88,ZAS91,FRO04}. 
Some interesting effects of dissipation on non-KAM chaos, including the identification of Cantor sets in phase space, were studied in Ref. \cite{VAS88}. But, to our knowledge, the bifurcation phenomena considered in the present paper, and their relation to resonance effects in SLs, have not been considered elsewhere. Moreover, what has not been clear previously is the nature of the characteristic instabilities and dynamical regimes that can be induced by dissipation in systems that exhibit non-KAM chaos, or how these regimes evolve with variation of the control parameters. Our analysis of such dynamics for electrons in a SL with a tilted magnetic field is therefore also of general interest in nonlinear dynamics and in the diverse areas of physics mentioned above, which involve the behavior of a harmonic oscillator driven by a plane wave.

\section{Effect of dissipation on chaotic dynamics.}

In Ref.\cite {FRO01}, the dynamics of a miniband electron in an electric and tilted magnetic field were studied in the absence of dissipation by solving Eq. (\ref{peneq}) with $\alpha=0$. Two distinct types of chaotic trajectories were identified from the different patterns that they produce in the Poincar\'e sections through phase space. The first type of pattern occurs only when the resonance condition 
\begin{eqnarray}  \label{ratio}
\omega_B=r\bar{\omega}_c
\end{eqnarray}
is satisfied, where $r$ is a rational number. At such resonances, the chaotic orbits map out a stochastic web comprising an intricate mesh of filaments, which threads the phase space and is a unique feature of non-KAM chaos. By contrast, the second type of pattern, a large continuous ``chaotic sea" in the Poincar\'e section, does not require the resonance condition to be satisfied. 
\begin{figure}[ht!]
\centering
\includegraphics[width=0.9\linewidth]{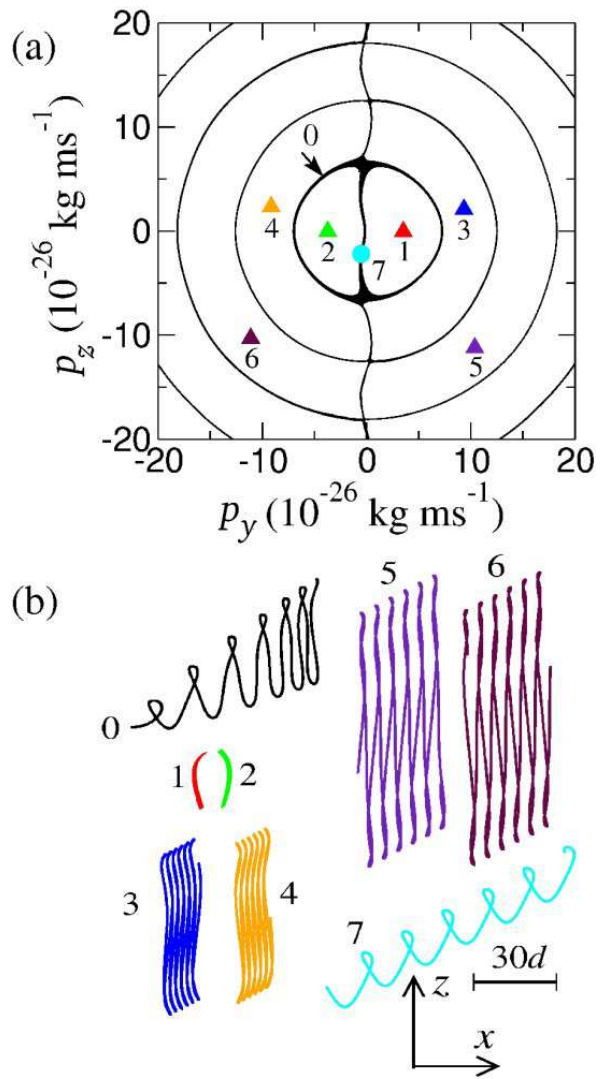}
\caption{(Color) (a) Black dots: stroboscopic Poincar\'e section calculated for the chaotic trajectory executed by an electron starting from rest in the dissipationless limit $\alpha=0$. This section reveals the inner filaments of a stochastic web. Colored triangles numbered 1-6: Poincar\'e sections calculated for six coexisting stable limit cycles formed when $\alpha= 10^{10}$ s$^{-1}$. Blue circle numbered 7: Poincar\'e section calculated for the single stable limit cycle found when $\alpha= 10^{12}$ s$^{-1}$. (b) Electron trajectories corresponding, by number and color, to the labeled features in the Poincar\'e sections shown in (a). All orbits are projected onto the $x-z$ plane (axes inset), have a common spatial scale shown by the horizontal line whose length is 30 SL periods, and are plotted over a fixed time interval of $9.45$ ps. The field parameters $B=2$ T, $\theta=30^{\circ}$ and $F=2.9\times10^5$ Vm$^{-1}$ satisfy the resonance condition $\omega_{B}/\bar{\omega}_{c}=1$.}
\label{web}
\end{figure}

We now illustrate these two types of chaos and investigate how the electron dynamics and associated phase space patterns change when dissipation is introduced by increasing $\alpha$ from zero in Eq. (\ref{peneq}). First, we consider the case when the phase space of the dissipation-free ($\alpha=0$) system contains a stochastic web corresponding to the $r = 1$ resonance condition attained when $F=2.9\times10^5$ Vm$^{-1}$, $B=2$ T, and $\theta=30^{\circ}$. The black dots in Fig. \ref{web}(a), which merge to form a continuous pattern, show a stroboscopic Poincar\'e section constructed from the electron trajectories by plotting the momentum components $(p_y,p_z)$ in the plane of the SL layers at discrete times separated by the Bloch period $T_B=2\pi/\omega_B$. 
The Poincar\'e section reveals a stochastic web, formed by the black dots and labeled ``0" in Fig. \ref{web}(a), 
which contains both ring-like and radial filaments. The radial filaments act as conduction channels, which enable the electron to diffuse rapidly outwards away from the web centre, thus increasing its momentum $p_L=(p_y^2+p_z^2)^{1/2}$ in the plane of the SL layers. In real space, the electron progresses rapidly along the SL axis \cite{foot2},
gaining kinetic energy from the electric field, which is transferred into the $y-z$ motion by the tilted magnetic field \cite{FRO01,FRO04}. To illustrate the resonant delocalization of the electron trajectories that results from stochastic web formation, the black curve labeled ``0" in Fig. \ref{web}(b) shows an electron orbit within the stochastic web projected onto the $x-z$ plane in which the magnetic field lies (axes inset). The electron starts from rest at the left-hand edge of the orbit, which corresponds the center point $(p_y,p_z)=(0,0)$ of the stochastic web in Fig. \ref{web}(a). A slow modulation of the density of orbital loops, which gradually increases with increasing $x$, is the only indication of irregularity in the electron trajectory labeled 0. For this reason, stochastic web chaos is often called ``weak chaos" \cite{ZAS91}, because the narrow width of the web filaments largely suppresses the highly erratic behavior usually found for nonintegrable systems. As the electron travels along the orbit from left to right in real space, it diffuses outwards along the almost vertical radial filament in the stochastic web towards higher $p_z$ values. The web filaments enmesh islands of stability, which are slices through invariant phase space tori generated by regular quasiperiodic trajectories \cite{FRO01,FRO04}. 
For clarity, in Fig. \ref{web}(a) we show no tori corresponding to orbits within these islands of stability.

The inclusion of even a very small dissipation term in Eq. (\ref{peneq}) destroys both the stochastic webs and islands of stability, replacing them with \emph{attracting} limit cycles to which all orbits tend as time progresses. To illustrate this, we first set $\alpha=10^{10}$ s$^{-1}$, which is far less than the momentum relaxation rate $W_m\simeq10^{12}-10^{13}$ s$^{-1}$ typical of electrons in semiconductor SL devices \cite{FRO01,FRO04}. For this value of $\alpha$, there are several distinct limit cycles, six of which are marked by the numbered and colored triangles in the stroboscopic Poincar\'e section shown in Fig. \ref{web}(a). Each limit cycle appears within, and attracts trajectories from, one of the islands of stability found when $\alpha=0$. Segments of the electron trajectories for each limit cycle are shown in Fig. \ref{web}(b). Each orbit is shown over the same fixed time interval of 9.45 ps, chosen when $t$ is large enough for the orbit to approach very close to the limit cycle, and linked, by number and color, to the corresponding points in the phase space plot shown in Fig. \ref{web}(a). The orbital segments reveal that 
the inclusion of weak dissipation makes the electron trajectories more regular because each limit cycle creates a particular repeating loop pattern.

As the level of dissipation increases, the different attracting limit cycles degenerate until only one remains in the phase space, whose position for $\alpha=10^{12}$ s$^{-1}$ is marked by the blue circle labeled ``7" in Fig. \ref{web}(a). Part of the corresponding electron orbit is shown by the blue curve labeled ``7" at the bottom of Fig. \ref{web}(b). The simple loop structure of this orbit transports the electron rapidly through the SL and is qualitatively similar to that of orbit 0, which generates the stochastic web in the absence of dissipation. So, although dissipation completely changes the phase space structure, replacing the stochastic web with a small number of limit cycles, it preserves the overall form of the delocalized orbits found when $\omega_B=r\bar{\omega}_c$ and actually \emph{enhances} resonant electron transport by preventing the spatial compression of the orbital loops, which slows the electron in the absence of dissipation (see, for example, orbit 0 in Fig. \ref{web}(b)).
\begin{figure}[t!]
\centering
\includegraphics[width=0.9\linewidth]{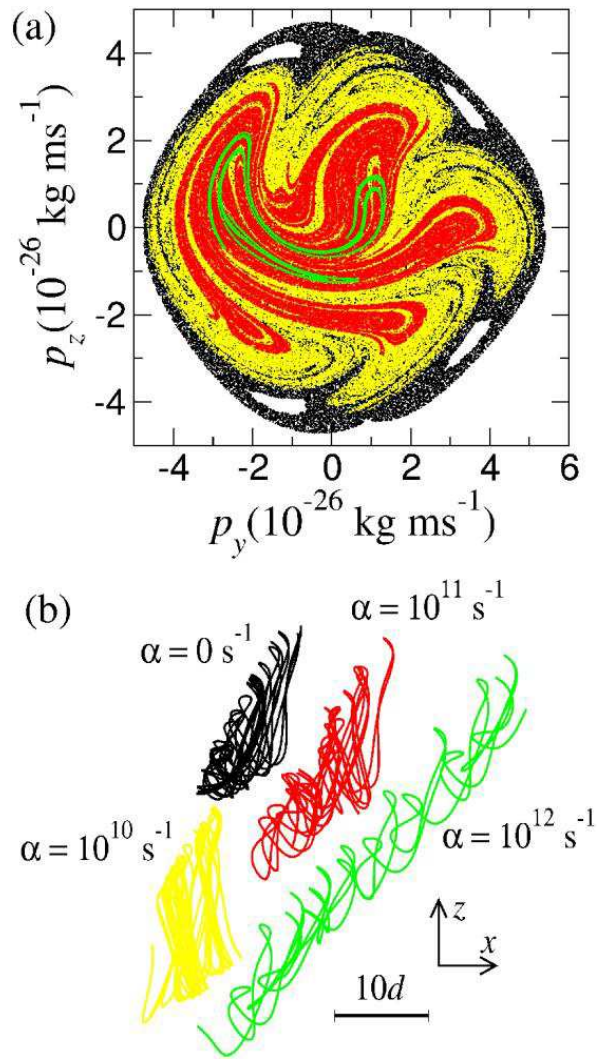}
\caption{(Color) (a) Black dots: stroboscopic Poincar\'e section calculated for an electron trajectory starting from rest in the dissipationless limit $\alpha=0$. Colored dots: stroboscopic Poincar\'e sections calculated for the chaotic attractors of electrons starting from rest when $\alpha= 10^{10}$ s$^{-1}$ (yellow dots),  $10^{11}$ s$^{-1}$, (red dots), and $10^{12}$ s$^{-1}$ (green dots). (b) Electron trajectories corresponding, by specified value of $\alpha$ and color, to the Poincar\'e sections shown in (a). All orbits are projected onto the $x-z$ plane (axes inset), have a common spatial scale shown by the horizontal line whose length is 10 SL periods, and are plotted over a fixed time interval of $22.68$ ps, which is long enough to reveal the form of the electron trajectories. $B=4.75$ T, $\theta=45^{\circ}$, $F=7.5\times10^5$ Vm$^{-1}$, $r\approx1.33$.}
\label{sea}
\end{figure}

We now consider the second type of Hamiltonian chaos, which, in the absence of dissipation, produces an extended chaotic sea in phase space even when $\omega_B/\bar{\omega}_c$ is not a rational number. The Poincar\'e section formed by the black dots in Fig. \ref{sea}(a) shows the chaotic sea generated by solving Eq. (\ref{peneq}) for an electron starting from rest in the SL with $\alpha=0$, $F=7.5\times10^5$ Vm$^{-1}$, $B=4.75$ T, and $\theta=45^{\circ}$, for which $r \approx1.33$.  Introducing dissipation into the equation of motion affects this type of extended chaotic sea far less than stochastic webs. In particular, although dissipation changes the form of the chaotic sea and the orbits within it, the orbits remain unstable and spatially irregular even for fairly large values of $\alpha$. The reason is that dissipation transforms the chaotic sea into a $chaotic$ attractor, rather than into a $stable$ limit cycle like those labeled 1-6 in Fig. 2. In Fig. \ref{sea}(b), the yellow, red and green dots show the chaotic attractors generated by plotting $(p_y,p_z)$ at every Bloch period for electron trajectories starting from rest with, respectively, $\alpha=10^{10}$ s$^{-1}$, $10^{11}$  s$^{-1}$, and $10^{12}$  s$^{-1}$. The irregular arrangement of points within these attractors demonstrates that they represent chaotic orbits, a conclusion that is confirmed by calculating their Lyapunov exponents. However, with increasing $\alpha$, the dimension of the chaotic attractors diminishes. For the three attractors shown in Fig.\ref{sea}(b), the Lyapunov dimension calculated using the Kaplan-Yorke formula \cite{KAP79, ANI01} is $2.987$, $2.887$, and $2.386$.
Chaotic electron orbits calculated for $\alpha=0$ (black curve), $10^{10}$  s$^{-1}$ (yellow curve),
$10^{11}$ s$^{-1}$ (red curve), and $10^{12}$ s$^{-1}$ (green curve) are shown projected into the $x-z$ plane in Fig. \ref{sea}(b). 
Comparison of the trajectories and Poincar\'e sections shown in Fig. \ref{sea} (a) and (b) for different values of $\alpha$ reveals that as the dissipation increases, the chaotic attractors gradually contract towards simple curves in phase space and, as a consequence, the orbits become spatially more regular. We emphasize that, in spite of the presence of energy dissipation, the electron trajectories remain chaotic, although their their topological structure in phase space changes as $\alpha$ varies over a wide range of values. In contrast to the Hamiltonian case, the phase space of the dissipative system contains unstable limit sets, which attract trajectories from certain regions of phase space known as basins of attraction. However, for very large $\alpha \gtrsim10^{13}$ s$^{-1}$ chaos completely disappears leaving only periodic orbits in the system.

\begin{figure}[t!]
\centering
\includegraphics[width=1.0\linewidth]{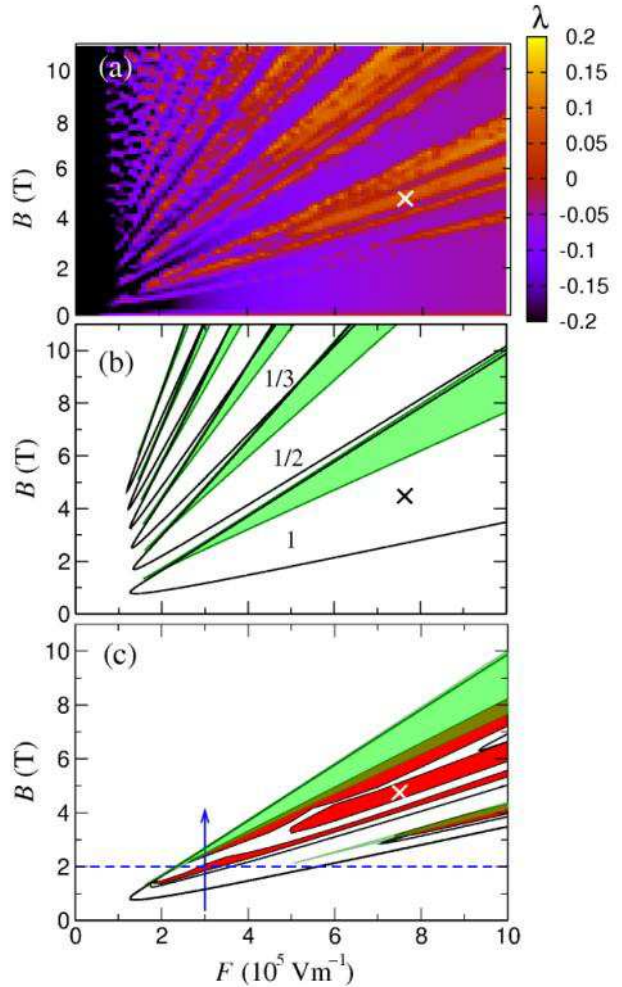}
\caption{(Color) (a) Color map showing $\lambda$ calculated as a function of $F$ and $B$ for the attractors approached by electron trajectories starting from rest. The color scale is shown on the right. (b) Black (green) curves are loci of the primary period-doubling (saddle-node) bifurcations of limit cycles corresponding to resonances with rational $r$ values (marked for $r=1,1/2,1/3$). Light green shaded areas are saddle-node tongues. (c) Location of chaotic attractors (red areas where $\lambda>0$) and additional bifurcations within the region of $F-B$ space bounded by the $r=1$ primary period-doubling bifurcation (outer black curve). Black (green) curves are loci of period-doubling and saddle-node bifurcations corresponding to rational values of $r\geqslant1$. Light green shaded areas are saddle-node tongues.
Dashed line marks $B=2$ T, and blue arrow shows the path through $F-B$ space corresponding to the bifurcation diagram in Fig. 5. In all panels, crosses mark $(F,B)$ coordinate corresponding to the chaotic orbit shown (green) in Fig. 3(b). $\theta=45^{\circ}$ and $\alpha=10^{12}$ s$^{-1}$.}
\label{bif1}
\end{figure}

\section{Stability and bifurcation analysis.}

In this section, we investigate how the stability of the electron orbits changes with the field parameters $F$, $B$, and $\theta$, and explore the bifurcation sequences that drive the transition between regular motion and dissipative deterministic chaos. We consider a fixed value of $\alpha=10^{12}$ s$^{-1}$.
First, we identify different regimes of electron dynamics by calculating the Lyapunov exponents for those limit sets that attract trajectories starting from rest.
To characterize the stability of the electron motion, we calculate the largest Lyapunov exponent, $\lambda$. The orbit is chaotic if $\lambda>0$, but otherwise is stable and regular.

The color map in Fig. \ref{bif1}(a) shows the value of $\lambda$ calculated as a function of $F$ and $B$ for $\theta=45^{\circ}$. For small $F$ or $B$, the system exhibits stable periodic oscillations, of frequency $\omega_B$, driven by the plane wave. We call such oscillations ``Period-1 limit cycles".  As $F$ or $B$ grows, the system undergoes a sequence of bifurcations, which eventually leads to the appearance of chaos within the yellow and red areas in Fig. \ref{bif1}(a), where $\lambda>0$. The crosses in Fig. \ref{bif1}(a-c) mark the coordinate in $F-B$ space for which an electron starting from rest eventually approaches the chaotic attractor shown green in Fig.\ref{sea}(b). 
The location and shape of the various islands of stable and chaotic electron motion in the $F-B$ plane [Fig. \ref{bif1}(a)] can be understood by considering the bifurcations that the limit cycles undergo as $F$ and/or $B$ increases. 
The black curves in Fig. \ref{bif1}(b) show the loci of the primary sequence of period-doubling bifurcations, i.e. those that occur first as $F$ increases. Each locus encloses a large ``V"-shaped region, which we call a ``period-doubling tongue" surrounding at least one area of chaos in the $F-B$ plane. When the field parameters enter one of these period-doubling tongues, the limit cycle electron trajectory undergoes a period-doubling bifurcation.
As a result of this bifurcation, the initial limit cycle loses its stability and, in its vicinity, another stable limit cycle appears whose period is twice as long.

A second type of bifurcation, namely the saddle-node bifurcation for limit cycles, also plays a key role in the electron dynamics. Loci of the primary saddle-node bifurcations are shown by the green curves in Fig. \ref{bif1}(b), which enclose ``saddle-node" tongues indicated by the green shaded areas. The top edges of these tongues are close to those of the period-doubling bifurcation loci [black curves in Fig. \ref{bif1}(b)] and lie near the lines $F=r(e\hbar\cos\theta/m^{*}ed)B$ ($r=1,1/2,1/3,...$) along which $\omega_B=r\bar{\omega}_c$. Consequently, each of the period-doubling and saddle-node tongues can be associated with a particular value of $r$ [those for $r=1,1/2,1/3,...$ are labeled in Fig. \ref{bif1}(b)], which relates to a resonance of a Period-1 cycle. Note, however, that when $\alpha\neq0$, nonlinear resonances can occur near, but not exactly at, field values for which $r$ is a rational number. When the field coordinate $(F,B)$ enters a saddle-node tongue, a new stable limit cycle is created, which corresponds to an unbounded electron trajectory, like that labeled ``7" in Fig. \ref{web}(a) for $r=1$. This new limit cycle is twinned with an unstable saddle counterpart. As the field coordinate $(F,B)$ leaves the boundary of the saddle-node tongue, the stable and saddle cycles approach one another in phase space and eventually combine via the saddle-node bifurcation that occurs at the edge of the tongue (we consider this process in more detail below).  At this bifurcation point, the stable and unstable saddle cycles both disappear, and the electron trajectory jumps abruptly onto another stable orbit located in a different part of phase space.

Within the area of $F-B$ space bounded by each of the primary period-doubling tongues in Fig. \ref{bif1}(b), the system undergoes a cascade of additional bifurcations. In Fig. \ref{bif1}(c), we illustrate this for the region of $F-B$ space enclosed by the $r=1$ primary period-doubling bifurcation (outer black curve). Within this region, further period-doubling and saddle-node bifurcations occur along the black and green curves respectively. For example, the sharp saddle-node and period-doubling tongues whose apexes are at $(F,B)\approx(5\times10^5$ Vm$^{-1}$, 2 T) 
in Fig. \ref{bif1}(c) are associated with the $r=2$ nonlinear resonance and produce orbits of period $2T_B$. Due to the presence of multiple tongues, the bifurcation diagram exhibits a complex self-similar structure. It is characterized by a cascade of bifurcations, which drives into chaos the electron trajectories that start from rest. These chaotic trajectories lie within the yellow and red areas of Fig. \ref{bif1}(a) and (c), where $\lambda>0$. 

An interesting effect occurs within those regions of $F-B$ space where period-doubling and saddle-node tongues intersect, for example the area marked by the cross in Fig. \ref{bif1}(c). As noted above, each period-doubling tongue contains a family of attractors. But in addition, the corresponding saddle-node tongue contains a \emph{distinct} stable limit cycle and a saddle (unstable) periodic orbit. Consequently, in the overlap region, at least two different attracting limit sets can coexist, which means that multistability occurs in the system. The initial condition of the orbit determines which attractor it eventually approaches.
\begin{figure}[th!]
\centering
\includegraphics[width=1.0\linewidth]{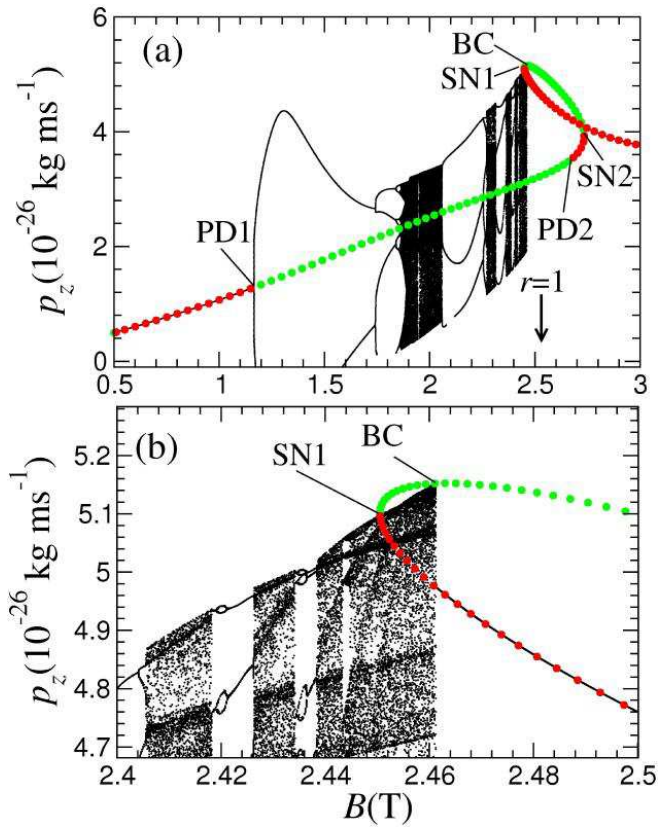}
\caption{(Color) (a) Poincar\'e sections constructed by plotting $p_z$ whenever $\dot{p}_z=0$ along limit cycles for electrons starting from rest (black dots) and for stable/unstable Period-1 limit cycles (red/green circles) with $F=3\times10^5$ Vm$^{-1}$, $\theta=45^{\circ}$ and $\alpha=10^{12}$ s$^{-1}$. Increasing $B$ induces period-doubling (PD1, PD2) and saddle-node (SN1, SN2) bifurcations, which correspond to crossing the black and green curves along the blue arrow in Fig. 4(c), and a boundary crisis (BC). Arrow marks $B$ value for which $r=1$. (b) An enlargement of the Poincar\'e section in the vicinity of saddle-node bifurcation SN1 and boundary crisis BC.}
\label{1dbif}
\end{figure}

To reveal in detail the bifurcations that the system undergoes inside the tongues in Fig. \ref{bif1}(c), we now consider how the electron orbits change as $B$ increases along the blue arrow in Fig. \ref{bif1}(c) with fixed $F=3\times10^5$ V m$^{-1}$. In particular, we construct the single-parameter bifurcation diagram shown in Fig. \ref{1dbif} by finding the limit sets that attract electrons starting from rest for $0.5\leq B\leq 3$ T and, for each $B$, plotting the values that $p_z$ attains (black dots in the figure) whenever $\dot{p}_z=0$ in the limit cycle. For $B\lesssim1.15$ T, this attractor is a Period-1 cycle, which produces a single black dot in Fig. \ref{1dbif}(a) at a $p_z$ value that gradually increases with increasing $B$. When $B\simeq1.15$ T, the Period-1 cycle undergoes a period-doubling bifurcation, labeled PD1 in Fig. \ref{1dbif}(a), which corresponds to entering the $r=1$ primary period-doubling tongue enclosed by the outer black curve in Fig. \ref{bif1}(c). As a result, $p_z$ takes one of two values when $\dot{p}_z=0$, causing the black dots to split into two distinct branches in Fig. \ref{1dbif}(a). As $B$ increases further, the limit cycle for electrons starting from rest undergoes a sequence of additional period-doubling bifurcations, leading to the onset of chaos when $B\simeq1.85$ T. Thereafter, chaotic motion alternates with ``stability windows", which is a characteristic feature of 
chaos induced by a cascade of period-doubling bifurcations \cite{ANI01}, and suddenly disappears when $B\approx2.461$ T, near the r=1 nonlinear resonance, which occurs at the $B$ value marked by the arrow in Fig. \ref{1dbif}(a). This suppression of chaos occurs because the attractor for electrons starting from rest switches back to a stable Period-1 cycle owing to a ``boundary crisis", which we explain in detail below. At this transition from chaotic to stable motion, the distribution of black dots in Fig. \ref{1dbif}(a), which is shown more clearly in the enlargement in Fig. \ref{1dbif}(b), changes from an irregular scatter to an ordered set of points located along a single-valued curve. 

To gain further insights into the electron dynamics, we now consider the nature and evolution of Period-1 cycles in the system over the entire range of $B$ values shown in Fig. \ref{1dbif}(a). To do this, we plot colored circles in Fig. \ref{1dbif} showing the values of $p_z$ whenever $\dot{p}_z=0$ in the Period-1 cycle. If the Period-1 cycle is stable (unstable) we plot a red (green) circle. For $B\lesssim1.15$ T, the Period-1 orbit is stable and is the only attractor in the system, being the limit cycle that all trajectories approach irrespective of their initial conditions.

 At the first period-doubling bifurcation [labeled PD1 in Fig. \ref{1dbif}(a)], the Period-1 cycle becomes unstable (and therefore no longer an attractor) and co-exists with the period-doubled stable limit cycle produced by the bifurcation. When $B$ reaches $\approx2.45$ T [labelled SN1 in Fig. \ref{1dbif}(a) and (b)] it enters the $r=1$ primary saddle-node tongue shown in Fig. \ref{bif1}(c). The associated saddle-node bifurcation creates \emph{new} stable and unstable saddle Period-1 cycles, which 
generate, respectively, the red and upper green circles in Fig. \ref{1dbif}(a),(b).

Thus, the two new Period-1 cycles formed by the saddle-node bifurcation coexist with the original Period-1 cycle, which produces the lower colored circles in Fig. \ref{1dbif}. This original Period-1 cycle stabilizes (circles change from green to red) when $B$ reaches $\approx2.67$ T and therefore crosses the upper edge of the $r=1$ primary period-doubling tongue shown in Fig. \ref{bif1}(c). At a slightly higher $B$ value of $\approx2.73$ T, it coalesces with the unstable saddle Period-1 cycle and thereby disappears as a result of saddle-node bifurcation SN2.
At this bifurcation, the field coordinate $(F,B)$ exits through the upper edge of the $r=1$ primary saddle-node tongue in Fig. \ref{bif1}(c).

At higher $B$ values there is a single Period-1 cycle, which is the only attractor in the system. 
Note that the nonlinear $r=1$ resonance has a pronounced effect on the bifurcation diagram in Fig. \ref{1dbif}. 
In particular, it gives rise to the characteristic loop formed by the different branches of the Period-1 limit cycles that coexist between bifurcations SN1 and SN2 in Fig. \ref{1dbif}(a) and also causes the abrupt disappearance of the chaotic attractor at the boundary crisis, BC. We now consider the latter in more detail.

\begin{figure}[h!]
\centering
\includegraphics[width=1.0\linewidth]{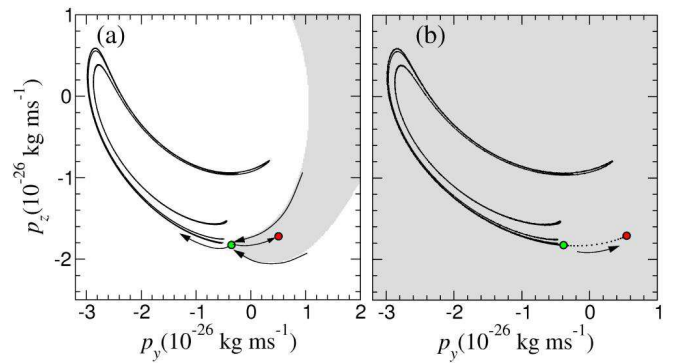}
\caption{(Color) Stroboscopic Poincar\'e sections calculated at times $t = lT_B$ $(l=0,1,2,...)$ when $B\approx$ (a) 2.46 T just before boundary crisis, (b) 2.461 T at the boundary crisis. The sections are constructed by plotting $(p_y, p_z)$ whenever $\dot{p}_z=0$ along the chaotic attractor for electrons starting from rest (black dots), and for the stable Period-1 limit cycle (red circles) and unstable saddle Period-1 limit cycle (green circle), which are formed by saddle-node bifurcation SN1 in Fig. 5. White (gray) regions are basins of attraction for the chaotic attractor and stable Period-1 limit cycle respectively. Arrows in (a) pointing towards (away from) the green circle represent stable (unstable) manifolds of the unstable saddle Period-1 limit cycle. Arrow and dotted curve in (b) show the direction along which the electron trajectory eventually leaves the chaotic attractor and subsequently approaches the stable Period-1 limit cycle. $F=3\times10^5$ Vm$^{-1}$, $\theta=45^{\circ}$ and $\alpha=10^{12}$ s$^{-1}$.}
\label{collision}
\end{figure}

As explained above, Fig. \ref{1dbif} reveals a small field range between $B\approx 2.450$ T and $B\approx 2.461$ T (i.e. between the points labeled SN1 and BC in the figure) where two distinct attractors coexist, a chaotic one and the stable Period-1 limit cycle. Figure \ref{collision}(a) shows a stroboscopic Poincar\'e section calculated for the chaotic attractor (black dots), stable Period-1 limit cycle (red circle), and unstable saddle Period-1 limit cycle (green circle) constructed by plotting the momentum components $(p_y,p_z)$ at times $t = lT_B$ $(l=0,1,2,...)$ when $B=2.46$ T, i.e. just before the boundary crisis that destroys the chaotic attractor.  Each attractor has a distinct basin of attraction comprising the set of initial conditions for which the trajectory will eventually approach the attractor. The basins of attraction for the chaotic set and Period-1 limit cycles are shown respectively by the white and gray areas of the $p_y-p_z$ plane in Fig. \ref{collision}(a). Note that the boundary between these two basins of attraction is the stable manifold of the unstable saddle periodic orbit (green circle),  which originates from the saddle-node bifurcation labeled SN1 in Fig. \ref{1dbif}. Trajectories with initial conditions exactly on this stable manifold approach the unstable saddle periodic orbit along the directions shown by the arrows pointing towards the green circle in Fig. \ref{collision}(a). Conversely, the unstable manifold of the saddle, represented schematically by the two arrows pointing away from  the green circle in Fig. \ref{collision}(a) repels electron trajectories away from the vicinity of the saddle orbit. Note that the lower right-hand part of the chaotic attractor in Fig. \ref{collision}(a) approaches very close to the boundary of its (white) basin of attraction. This suggests that if changing $B$ increases the size of the chaotic attractor, there will be a boundary crisis, which occurs when an attractor hits the boundary of its basin of attraction. 

To investigate this possibility, we studied how the phase space structure shown in Fig. \ref{collision}(a) responds to small changes of $B$. Our calculations confirm that increasing $B$ to $\approx 2.461$ T does indeed trigger a boundary crisis, which causes the abrupt disappearance of the chaotic attractor at the field value labeled BC in Fig. \ref{1dbif}. To illustrate this, Fig. \ref{collision}(b) shows the stroboscopic Poincar\'e section calculated for the chaotic set (black dots), stable Period-1 limit cycle (red circle) and unstable saddle Period-1 limit cycle (green circle) at the boundary crisis. The lower right-hand edge of the chaotic attractor touches the saddle (green circle), which, for lower $B$, was located on the boundary between the basins of attraction of the chaotic and Period-1 attractors [Fig. \ref{collision}(a)]. When the chaotic attractor and saddle point touch at the boundary crisis, all trajectories that approach the chaotic attractor eventually leave it at the bottom right-hand tip in Fig. \ref{collision}(b), then cross the curve in the $p_y-p_z$ plane that used to separate the two basins of attraction, and thereafter move in the direction of the arrow along the dotted curve in Fig. \ref{collision}(b) towards the stable Period-1 limit cycle (red circle). Consequently, for \emph{all} initial conditions, the electron trajectory eventually approaches the stable Period-1 limit cycle, either directly or via the tip of the former chaotic attractor, which now allows the electron trajectories to escape. Over the field range between the boundary crisis and the second period-doubling transition (i.e. between the fields labeled BC and PD2 in Fig. \ref{1dbif}) the Period-1 limit cycle is therefore the only attractor in the system and its basin of attraction [gray in Fig. \ref{collision}(b)] fills the entire phase space.

We now consider how the attractors evolve as $B$ sweeps either up or down through the boundary crisis. As noted above, for $B$ values in the range $\approx 1.15$ to $\approx 2.45$ T (i.e. between points PD1 and SN1 in Fig. \ref{1dbif}), the chaotic set (black dots in Fig. \ref{1dbif}) is the only attractor in the system. As $B$ increases above $\approx 2.467$ T, the state of the system remains pinned to the chaotic attractor even though a second  attractor (red circles in upper part of Fig. \ref{1dbif}) emerges from the saddle-node bifurcation labeled SN1 in Fig. \ref{1dbif}. When $B$ reaches the value at which the boundary crisis occurs, the system state changes abruptly from chaotic to the  Period-1 attractor, as discussed above. Conversely, if $B$ now decreases, the state remains pinned to the  Period-1 attractor until the saddle-node bifurcation SN1 [where the red and green circles meet in Fig. \ref{1dbif}(b)] removes this attractor when $B\approx 2.45$ T, i.e. at a field value \emph{below} that corresponding to the boundary crisis. Consequently, the transition between chaotic attractors and regular limit cycles occurs at a higher $B$ value when the field sweeps up through the boundary crisis than when it sweeps down, and the system exhibits {\it hysteresis}. For given $F$, the hysteresis occurs over the magnetic field range for which the $r=1$ saddle-node tongue, shown by the upper gray region in Fig. \ref{bif1}(c), overlaps with the upper red island in the figure, in which the attractor is chaotic. Figure \ref{bif1}(c) shows that this overlap region broadens with increasing $F$, meaning that the hysteresis becomes more pronounced, that is, it extends over a wider magnetic field range. The occurrence of hysteresis in bifurcation diagrams like Fig. \ref{1dbif} is a common feature of resonances in non-linear systems \cite{GUC93}.

Note that two distinct mechanisms cause the onset of chaotic limit cycles in the system. As $B$ increases from 0, the transition to chaos occurs via the cascade of period-doubling bifurcations shown in Fig. \ref{1dbif}. Conversely, as $B$ decreases from 3 T, chaos switches on abruptly at the saddle-node bifurcation (point SN1 in Fig. \ref{1dbif}) that removes the stable Period-1 limit cycle.

\begin{figure}[th]
\centering
\includegraphics[width=1.0\linewidth]{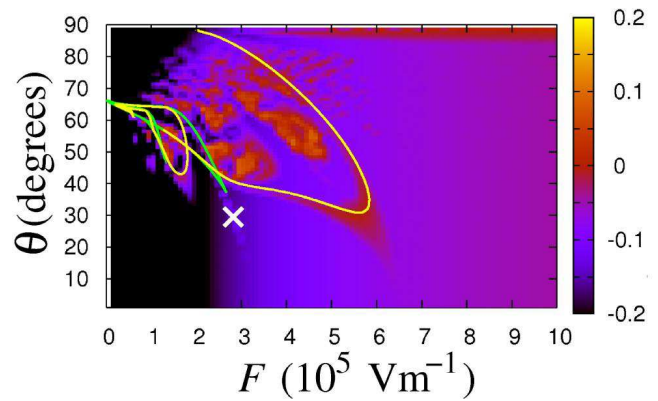}
\caption{(Color) (a) Color map showing $\lambda$ calculated as a function of $F$ and $\theta$ for the attractors approached by electron trajectories starting from rest when $B=2$ T and $\alpha=10^{12}$ s$^{-1}$. Color scale is shown on the right. Yellow (green) curves are loci of period-doubling (saddle-node) bifurcations of electron limit cycles. Cross marks $(F,\theta)$ coordinate corresponding to limit cycle 7 in Fig. 2.}
\label{bif2}
\end{figure}

The two main mechanisms responsible for the onset of chaos (the period-doubling sequence and saddle-node bifurcation) occur for a wide range of $\theta$ values. This can be seen from Fig. \ref{bif2}, which shows the loci of the main period-doubling (yellow curves) and saddle-node (green curves) bifurcations in the $F-\theta$ plane calculated for fixed $B=2 T$. These loci are overlaid on a color map showing the value of the largest Lyapunov exponent for the limit cycles approached by electron trajectories starting from rest. Note that, as in Fig. \ref{bif1}, the limit cycles become chaotic (yellow or red in the color map) within the regions of $F-\theta$ space bounded by the period-doubling bifurcations (yellow lines), which partially overlap the areas enclosed by the saddle-node bifurcations (green lines). As a consequence of this overlap, multistability, and the associated hysteretic evolution of limit cycles, occurs for a range of $F$ and $\theta$ in the same way that hysteresis occurs within islands in the $F-B$ plane (see  Fig. \ref{bif1}). Note, however, that for the particular field value $B=2$ T used to calculate Fig. \ref{bif2}, chaos only occurs for $\theta>40^{\circ}$. For smaller values of $\theta$, electrons starting from rest approach only stable limit cycles, for example that labeled 7 in Fig. 2, which is calculated for the $(F,\theta)$ coordinate marked by the cross in Fig. \ref{bif2}.
\begin{figure}[th]
\centering
\includegraphics[width=.9\linewidth]{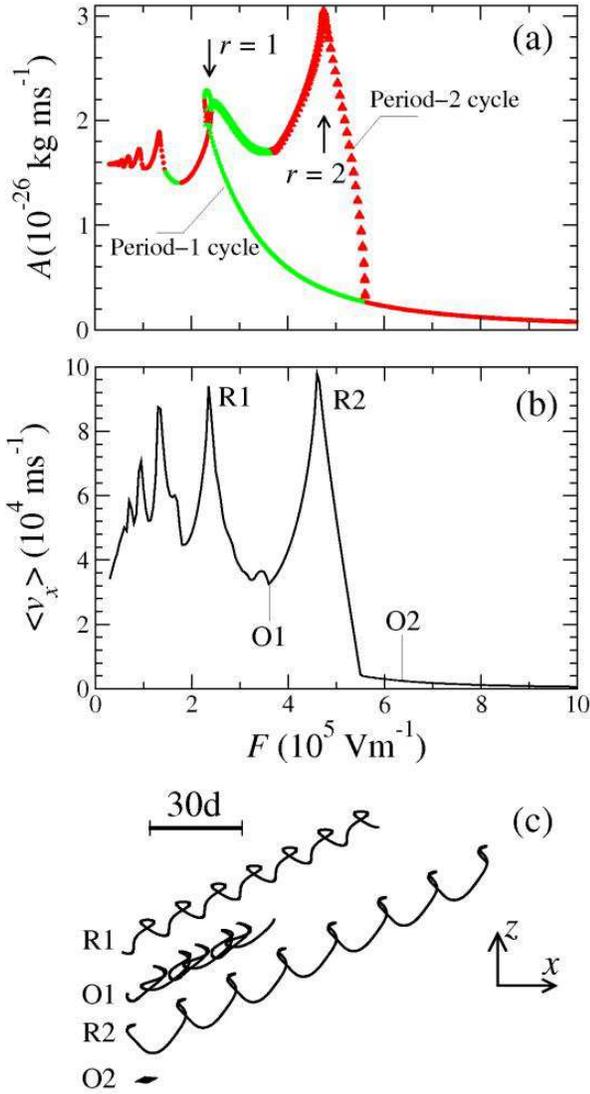}
\caption{(Color) (a) Amplitude, $A$, of $p_{z}(t)$ oscillations calculated versus $F$ for a stable/unstable Period-1 limit cycle (red/green circles) and a stable/unstable Period-2 limit cycle (red/green triangles). Arrows mark positions of $r$ = 1 and 2 resonances. (b) $\langle v_x\rangle$ calculated as a function of $F$ for the attractor approached by electrons starting from rest. Peaks labeled R1(R2) occur at $F$ values near the $r=1(2)$ nonlinear resonances. 
Labels O1(O2) mark off-resonance field values for which $r=1.52(2.53)$. (c) 
Orbits approached by electrons starting from rest at the resonant (R1,R2) and off-resonant (O1,O2) $F$ values marked in (b). All attractors are projected onto the $x-z$ plane (axes inset), have a common spatial scale shown by the horizontal line whose length is 30 SL periods, and are plotted over the same time interval.
$B=2 T$, $\theta=45^{\circ}$, $\alpha=10^{12}$ s$^{-1}$.}
\label{res}
\end{figure}

\section{The effect of resonances on electron velocity}

In Section II, we showed that electron motion in the SL can be described by Eq. (\ref{peneq}), corresponding to a damped harmonic oscillator driven by a plane wave. When the driving frequency, $\omega_{B}\propto{F}$, is commensurate with the natural frequency, $\bar{\omega_{c}}\propto{B}$, of the harmonic oscillator nonlinear resonances occur in the system. The effect of such resonances on the electron dynamics is complex. For example, in Sections III and IV we saw that the $r=1$ resonance creates a stable limit cycle, which corresponds to a spatially unbounded electron trajectory, via a saddle-node bifurcation, and also triggers a boundary crisis. In this section, we investigate how resonances affect the momentum and drift velocity of the electrons, both parallel and perpendicular to the SL layers. We focus on these dynamical variables because they strongly influence the transport characteristics, such as $I(V)$ curves, measured in experiments \cite{FRO04}.

First, we consider how resonances affect the variation of the in-plane momentum component, $p_z$, as the electron performs a limit cycle. This momentum component is the most natural parameter to study because it is the dependent variable in Eq. (\ref{peneq}). For each limit cycle, we define the amplitude of the $p_z$ oscillations to be $A=(p_z^{max}-p_z^{min})/2$, where $p_z^{min}$ and $p_z^{max}$ are, respectively, the minimum and maximum values of $p_z$ attained during the limit cycle. We then investigate how $A$ varies with $F$ for several different limit cycles, including those produced by the bifurcations induced by changing $F$. In Fig. \ref{res}(a), we show $A$ calculated as a function of $F$ for Period-1 (circles) and Period-2 (triangles) limit cycles when $B=2$ T and $\theta=45^{\circ}$. The $A(F)$ plots, analogous to the frequency response curves familiar to engineers, exhibit the resonant features that characterize nonlinear dissipative systems. The resonances are clearly seen as sharp increases of $A$ when $r$ is close to rational \cite{footnote2}, with the largest peaks [arrowed in Fig. \ref{res}(a)] occurring near the $F$ values for which $r=1$ and 2. The smaller peaks seen at lower $F$ correspond to rational $r<1$. A Period-1 limit cycle exists over the entire range of $F$ shown in Fig. \ref{res}(a), but undergoes bifurcations as $F$ increases, which changes it between stable (red circles) and unstable (green circles). The location of the primary bifurcations for $B=2$ T and $\theta=45^{\circ}$ can be seen from Fig. \ref{bif2}. For $F$ values close to $2.35\times10^{5}$ Vm$^{-1}$, when $r\approx1$, there is a saddle-node tongue (enclosed by the green curve in Fig. \ref{bif2}), which means that up to three Period-1 orbits coexist (as in the loop region of Fig. \ref{1dbif}). Consequently, the $A(F)$ curve for the Period-1 limit cycle contains a loop when $r\approx1$. Within the loop region, the stable Period-1 limit cycle undergoes a period-doubling bifurcation (along the right-hand yellow curve in Fig. \ref{bif2}), which produces the Period-2 limit cycle whose $A(F)$ curve [triangles in Fig. \ref{res}(a)] reveals large resonant peaks when $r=1$ and 2. 

\begin{figure}[t]
\centering
\includegraphics[width=.9\linewidth]{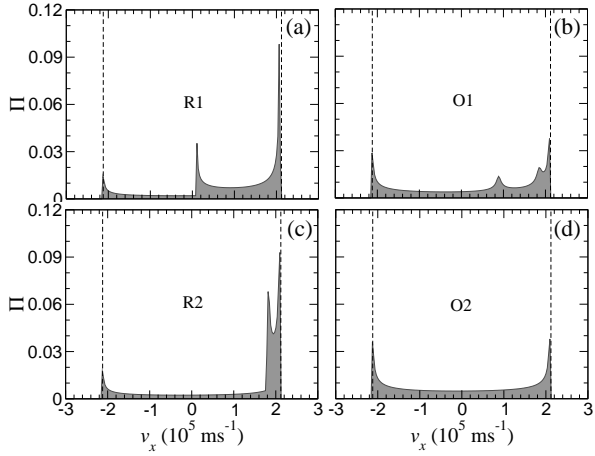}
\caption{Normalized $\Pi(v_x)$ curves (with gray underfill) calculated for electron trajectories labeled in Fig. \ref{res} (b) and (c) as: (a) R1, (b) O1, (c) R2, (d) O2. Vertical dashed lines mark extremal velocities $v_x=\pm{v_{max}}$. $B=2$ T, $\theta=45^{\circ}$, $\alpha=10^{12}$ s$^{-1}$.}
\label{vx}
\end{figure}

The electron velocity component, $v_x$, along the SL axis is a particularly important dynamical variable because it determines the electrical current that flows through the device and the power dissipation rate per particle, $P_d=eFv_x$. It follows from the equations of motion (\ref{peneq}) and (\ref{veleq}) that for each attractor, $v_x$ oscillates as a function of time. To determine the drift component of the electron velocity, which determines the current flow \cite{FRO04}, we therefore calculate the time average $\langle v_x \rangle$ of the velocity throughout the attractor. Figure \ref{res}(b) shows $\langle v_x\rangle$ calculated as a function of $F$ for the attractor approached by electrons starting from rest with $B=2$ T and $\theta=45^{\circ}$. The shape of this graph is similar to the $A(F)$ data shown in Fig. \ref{res}(a). In particular, there is strong resonant enhancement of $\langle v_x\rangle$ near rational $r$ values where maxima also occur in the $A(F)$ curves calculated for the Period-1 and Period-2 limit cycles. The peak labeled R1 in the $\langle v_x\rangle$ versus $F$ curve corresponds to the $r=1$ resonance of the Period-1 limit cycle, whereas peak R2 originates from the $r=2$ resonance of the Period-2 limit cycle. Those peaks to the left of R1 arise from resonances of the Period-1 limit cycle. To relate the resonant peaks in the $\langle v_x\rangle$ versus $F$ curve directly to the electron motion, Fig. \ref{res}(c) shows segments of Period-1 (lower and two upper trajectories) and Period-2 (third orbit down) limit cycles calculated at the field values labeled R1, R2, O1, and O2 in Fig. \ref{res}(b) over a fixed time interval of 12.38 ps. During this time interval, the electron travels much further along in the two orbits on resonance (R1 and R2) than in the two off-resonance orbits (O1 and O2). 
\begin{figure}[t]
\centering
\includegraphics[width=.9\linewidth]{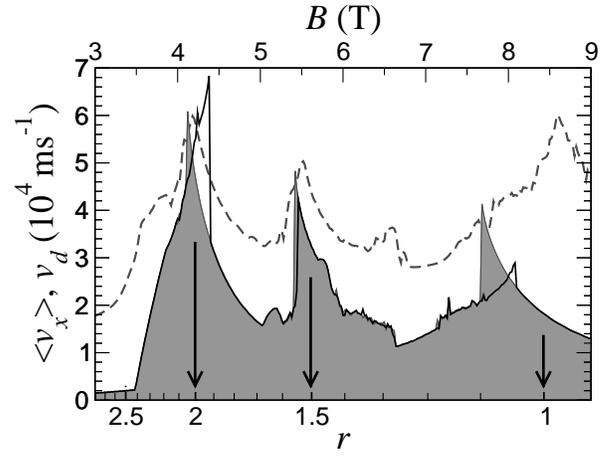}
\caption{$\langle v_x\rangle$ calculated as a function of increasing (black curve) and decreasing (under-filled gray curve) $B$ (upper axis). Lower axis shows values of $r\propto 1/B$. Peaks in $\langle v_x\rangle$ originate from $r=1,3/2,$ and 2 resonances (located by arrows). The difference between the two curves originates from the hysteretic evolution of the electron limit cycle discussed in Section IV. Dashed curve shows $v_d$ (with \emph{no} vertical offset) calculated as a function of $B$ (or $r$) using Eq. (\ref{eq:vd}). $F=10^{6}$ Vm$^{-1}$, $\theta=45^{\circ}$, $\alpha=10^{12}$ s$^{-1}$.}
\label{hyst}
\end{figure}

To further investigate how resonances affect the electron transport, we calculated the probability distribution, $\Pi(v_x)$, of all the $v_x$ values attained during one period of the limit cycles approached by electrons starting from rest both on and off resonance. Figure \ref{vx} shows normalized $\Pi(v_x)$ curves calculated for the orbits labeled (a) R1, (b) O1, (c) R2, (d) O2 in Fig. \ref{res} (b) and (c).
All attainable $v_x$ values lie between the two vertical dashed lines located at $v_x=\pm{v_{max}}$, where $v_{max}=d\Delta/2\hbar$ is the maximum velocity of an electron in the lowest miniband. Comparison of the velocity 
{\b distributions on (a),(c) and off (b),(d) resonance} reveals that on resonance, the most probable $v_x$ value approaches $v_{max}$, thus producing the large peak $\langle v_x\rangle$ values shown in Fig. \ref{res}(b).

The hysteretic response exhibited by the limit cycles as $B$ is swept up and down through a boundary crisis has a clear manifestation in the $\langle v_x\rangle$ versus $B$ curves calculated for large $F$ values where the hysteresis is most pronounced, as discussed in Section IV. To illustrate this, Fig. \ref{hyst} shows $\langle v_x\rangle$ calculated as a function of increasing (black curve) and decreasing (under-filled gray curve) $B$ (upper axis) when $F=10^{6}$ Vm$^{-1}$ and $\theta=45^{\circ}$. Note that the lower axis of the graph shows the ratio, $r\propto 1/B$, of the Bloch and cyclotron frequencies.  Near the $r=1, 3/2,$ and $2$ resonances (marked by arrows in Fig. \ref{hyst}), saddle-node tongues coexist with chaotic attractors. This can be seen, for example, from Fig. \ref{bif1}(c), which, as discussed in Section IV, reveals a large overlap between the $r=1$ saddle-node tongue and upper red island of chaos when $F=10^{6}$ Vm$^{-1}$. As a consequence, the limit cycle evolves in a different way when $B$ increases or decreases, and so the corresponding $\langle v_x\rangle$ versus $B$ curves exhibit hysteresis. This hysteresis is particularly pronounced near the $r=1$ and $r=2$ resonances in Fig. \ref{hyst}, which reveals that for some $B$ values $\langle v_x\rangle$ differs by a factor of two between up and down sweeps of $B$. Since the current flow depends critically on the electron drift velocity, we predict that measured $I(B)$ curves should also reveal strong resonant peaks that exhibit clear hysteresis.

In Fig. \ref{hyst}, we compare the $\langle v_x\rangle$ versus $B$ curves obtained by including dissipation \emph{a priori} in the equations of motion, with drift velocities, $v_d(B)$, calculated using the traditional Esaki-Tsu approach \cite{ESA70}, based on non-dissipative electron trajectories. Within this approach, which we used in our previous studies of electron transport in SLs with a tilted magnetic field \cite{FRO01,FRO04,Trav,STA04}, 
\begin{equation}
v_d=\frac{1}{\tau}\int\limits^{\infty}_0 \exp\left(\frac{-t}{\tau}\right) v_x (t) dt,
\label{eq:vd}
\end{equation}
where $v_x (t)=\dot{x}$ is determined from Eqs. (\ref{peneq}) and (8), setting $\alpha=0$, and the electron scattering time, $\tau$, includes contributions from both elastic and inelastic scattering processes \cite{FRO04}. The $v_d(B)$ variation obtained from Eq. (\ref{eq:vd}) is shown by the dashed curve in Fig.~ \ref{hyst}. Comparison with the $\langle v_x\rangle$ versus $B$ curves (black and under-filled gray) reveals that both approaches give qualitatively similar results, with prominent maxima at $B$ values close to $r=$ 1, 3/2, and 2. However, the positions of these maxima do not always coincide exactly, in particular for $r=1$, mainly because including dissipation \emph{a priori} in the equations of motion shifts the resonances slightly away from rational values of $r$, as discussed in Section IV. Although the resonant peaks in the $\langle v_x\rangle$ versus $B$ and $v_d(B)$ curves have very similar heights, especially for $r$ = 2 and 3/2, away from the resonances $\langle v_x\rangle$ is significantly lower than $v_d$. The reason for this is that off resonance, $p_z$ increases less rapidly with increasing $t$, which means that the damping term, $\alpha\dot{p}_z$, in Eq. (\ref{peneq}) remains small and therefore causes less scattering-induced transport.

Figure \ref{hyst} reveals that, for the SL considered here, the most pronounced resonant enhancement of $\langle v_x\rangle$ occurs near $r=2$ [as also shown in Fig. \ref{res}(b)]. The two other resonances shown in Fig.~\ref{hyst} near $r=$ 1 and 3/2 weaken as $r$ decreases because the corresponding increase of $B$ reduces the distance that the electrons travel along the $x$ axis in a given time interval [compare, for example, orbits R2 and R1 calculated for $r$ = 2 and 1 respectively in Fig. \ref{res}(c)]. 

\begin{figure}[t]
\centering
\includegraphics[width=.9\linewidth]{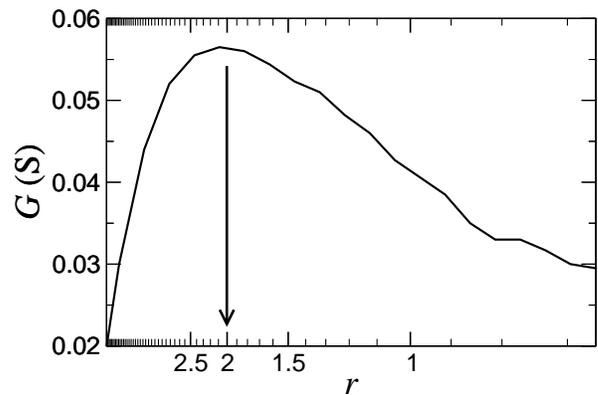}
\caption{Experimental values of the low-temperature (4.2 K) differential conductance, $G=dI/dV$, versus $r=\omega_B/\bar{\omega}_c\propto F/B$, measured for different values of $B$ at fixed $V$ = 0.25 V and $\theta=45^{\circ}$ for the SL described in Ref. \cite{FRO04}, whose parameters are similar to those specified in Section~\ref{sec:model}. Arrow marks $r=2$ resonance.} 
\label{exp}
\end{figure}

To investigate how such resonances manifest themselves in the electrical properties of a real device, we measured $I(V)$ characteristics for the SL studied in Ref. \cite{FRO04}, whose parameters are similar to those described in Section~\ref{sec:model}. Figure~\ref{exp} shows the differential conductance $G$=$dI/dV$ measured as function of $r$ at $V$ = 0.25 V and $\theta=45^{\circ}$. In the experiments, we varied $B$ and, for each value, measured $I$ over a very small range of $V$ near 0.25 V, taking the derivative to find $G$. For each $B$ value, we determined the corresponding $r \propto F/B$ by assuming that $F \propto V$ and is uniform throughout the SL. The dominant feature in Fig. \ref{exp} is the pronounced conductance peak at the $r=2$ resonance (arrowed). Additional, far weaker, features can be seen for smaller $r$, but are not strong enough for us to link them definitively to particular $r$ values. The overall shape of the experimental $G(r)$ curve in Fig. \ref{exp} is broadly similar to the $\langle v_x\rangle$ versus $r$ plots shown in Fig.~\ref{hyst}. However, only the $r$ = 2 conductance peak is clearly visible in the experimental data. This difference between experiment and theory occurs because in the actual device, $F$ is not constant throughout the device, as assumed in our present calculations, but varies with $x$ due to the formation of charge domains \cite {FRO04}. The spatial variation of $F$ means that the resonance condition is only satisfied in a small region of the SL, which weakens and blurs the resonances in $G(r)$ \cite{Trav}. This effect is less pronounced for stronger resonances, at which the electric field is more uniform through the SL \cite{Trav}. Consequently, the $r$ = 2 resonance is clearly revealed and correctly positioned in Fig. \ref{exp}. 
  
\section{Summary and Conclusions.}

We have shown that both Hamiltonian and dissipative chaos strongly affect the transport of miniband electrons in biased SLs with
a tilted magnetic field. Chaos originates from a complex nonlinear interaction between Bloch and cyclotron oscillations associated, respectively, with the electric field and magnetic field component, $B\cos\theta$, along the SL axis. In the absence of dissipation, the system exhibits non-KAM chaos, characterized by the formation of intricate stochastic web patterns when $\omega_B$ and $\bar{\omega}_c$ are commensurate. When weak dissipation is included \emph{a priori} in the equations of motion, limit cycles replace the stochastic web filaments and stable orbits that they enclose. As the dissipation strength increases, the number of limit cycles and the volume of phase space occupied by chaotic trajectories both decrease. However, for the electron momentum relaxation rates measured for SLs used in recent experiments \cite{FRO01,FRO04}, chaos occurs for a wide range of $F$ and $B$ values, usually when $\theta> 40^{\circ}$ (see Figs. \ref{bif1}, \ref{bif2}). 

Two distinct bifurcation mechanisms drive the transition to chaos: a period-doubling cascade, where the onset of chaos is gradual, and a boundary crisis, which switches chaos on abruptly. Both mechanisms operate for all tilt angles in the range $0<\theta<90^{\circ}$. For both Hamiltonian and dissipative models of the electron dynamics, the onset of chaos occurs near resonances when $\omega_B$ and $\bar{\omega}_c$ are commensurate. 
Such resonances cause an abrupt delocalization of the electron trajectories, due either to stochastic web formation in a Hamiltonian picture, or to the creation of attractors corresponding to unbounded stable limit cycles when dissipation is included \emph{ab initio} in the equations of motion. This delocalization produces large resonant peaks in the electron drift velocity, which, in turn, generate the strong resonant enhancement of the current flow observed in our recent experiments \cite{FRO04}. 
In the case of dissipative dynamics, the drift velocity increases near resonances because the most probable value of $v_x$ that the electron attains during its unbounded limit cycle is very close to the maximal possible value, determined by the SL parameters (Fig. \ref{vx}). Resonances between $\omega_B$ and $\bar{\omega}_c$ also give rise to multistability phenomena when two or more different attractors coexist in phase space. Multistability makes limit cycles, and the corresponding $\langle v_x\rangle$ values, evolve differently when $F$ or $B$ sweeps up or down through a resonance. We expect that this hysteresis will also have striking experimental signatures. In particular, our calculations suggest that experimental $I(B)$ or $I(V)$ curves will exhibit pronounced hysteresis, reflecting that shown in Fig. \ref{hyst} for electron drift velocities, especially at high field values. 

Finally, we emphasize that the driven harmonic oscillator equation (\ref{peneq}), which describes the motion of miniband electrons in a tilted magnetic field, is also of fundamental importance in many other physical systems, for example plasma \cite{SAG88}, ultra-cold atoms in optical lattices \cite{HEN01,STE01,SCO02}, the transmission of light through photonic crystals \cite{WIL03}, turbulent flow and in understanding the patterns of quasi-crystals \cite{ZAS91}. Although the motion of the \emph{dissipation-free} harmonic oscillator driven by a plane wave is well understood, to our knowledge the effects of dissipation have not previously been considered. Our study of the rich dynamics of the \emph{dissipative} driven harmonic oscillator is relevant to a wide range of topics in physics, engineering, and applied mathematics.

\section*{ACKNOWLEDGEMENTS}

The work was supported by the UK Engineering and Physical Sciences Research Council. We thank M. Henini (University of Nottingham) for MBE growth of the SL samples used in our experiments, R. Airey (University of Sheffield) for processing the samples, and D.K. Maude (Grenoble High Magnetic Field Laboratory) for assistance with the high magnetic field measurements. 


\begin{thebibliography}{99}
\bibitem{footnote2} In contrast to the conservative ($\alpha=0$) case, when dissipation is included in Eq. (6), the resonances usually occur near, but not exactly at, rational $r$ values.
\end{thebibliography}

\end{document}